\documentclass[pra,aps,groupedaddress,floatfix,twocolumn,superscriptaddress,showpacs]{revtex4-1}
\usepackage{amsmath,amssymb,multirow,epsfig,bm,pifont}
\usepackage[linkcolor=black,citecolor=black,urlcolor=blue,colorlinks=true,linktocpage=true]{hyperref}
\usepackage{graphicx}
\usepackage{epstopdf}
\usepackage{bm}
\usepackage{tabularx}
\usepackage{makecell}

\newcommand{\etal}{{\it et al.}}
\newcommand{\beq}{\begin{equation}}
\newcommand{\eeq}{\end{equation}}
\newcommand{\bea}{\begin{eqnarray}}
\newcommand{\eea}{\end{eqnarray}}

\newcommand{\bd}{\boldsymbol}

\begin{document}
\title{Ground state of the two-dimensional attractive Fermi gas: \\ essential properties from few- to many-body}

\author{Lukas Rammelm\"uller}
\email{lukas.rammelmueller@tuwien.ac.at}
\affiliation{Institute of Solid State Physics, Vienna University of Technology, A-1040 Vienna, Austria}
\affiliation{Department of Physics and Astronomy, University of North Carolina, Chapel Hill, NC, 27599, USA}

\author{William J. Porter}
\email{wjporter@live.unc.edu}

\author{Joaqu\'{\i}n E. Drut}
\email{drut@email.unc.edu}
\affiliation{Department of Physics and Astronomy, University of North Carolina, Chapel Hill, NC, 27599, USA}

\begin{abstract}
We calculate the ground-state properties of unpolarized two-dimensional attractive fermions in 
the range from few to many particles. Using first-principles lattice Monte Carlo methods,
we determine the ground-state energy, Tan's contact, momentum distribution, and 
single-particle correlation function. We investigate those properties for systems of $N=4,8,...,40$ 
particles and for a wide range of attractive couplings. As the attractive coupling is increased,
the thermodynamic limit is reached at progressively lower $N$ due to the dominance of the two-body
sector. At large momenta $k$, the momentum distribution displays the expected $k^{-4}$ behavior, but
its onset shifts from $k \simeq 1.8 k^{}_F$ at weak coupling towards higher $k$ at strong coupling.
\end{abstract}

\date{\today}
\maketitle

\section{Introduction}
Precise experiments with ultracold atomic fermion clouds
are currently being carried out by several groups around the world.
Among the systems in the ever-expanding set that such experiments can study,
it is now possible to probe two-dimensional (2D) physics in a clean and controllable way~\cite{RevExp,
RevTheory,Parish2DReview}. This is an exciting opportunity to understand
key aspects of few- and many-body quantum physics that are not specific to
atoms but which are generic to 2D quantum mechanics.

Such phenomena include the classical scale 
invariance displayed by non-relativistic fermions in 2D, which is broken by quantum fluctuations 
(i.e. the symmetry is anomalous, see~\cite{ScaleInvariance}), a property shared with
four-dimensional gauge theories like quantum chromodynamics~\cite{QCD}. 
Although finite-temperature symmetry-breaking transitions are not
possible for continuous symmetries in 2D (as explained by the Mermin-Wagner theorem~\cite{MW}), 
attractive interactions do result in a Berezinskii-Kosterlitz-Thouless (BKT) transition into a low-temperature superfluid phase~\cite{BKT},
which is another feature generic to 2D systems. 
High-temperature superconductivity is also understood to be essentially a 2D phenomenon~\cite{HighTc}, and it shares with ultracold
atoms the so-called pseudogap regime~\cite{pseudogap}. Last, but not least, the recent excitement about the physics
of graphene is also associated with scale invariant 2D systems, and with its affinity with relativistic strongly coupled 
matter~\cite{graphene}.

Thus, the realization and exploration of flat ultracold atomic
clouds impacts a wide range of areas in physics, and in the last few years this research has been pursued vigorously
(see Refs.~\cite{RanderiaPairingFlatLand, PieriDanceInDisk} for recent wide-audience reports).
On the experimental side, two-dimensional fermionic clouds were first achieved just a few years ago
in~\cite{Experiments2D2010Observation, Experiments2D2011Observation}, and many properties, ranging from
spectroscopy to thermodynamics and hydrodynamic response, have been 
studied since~\cite{Experiments2D2011RfSpectroscopy, Experiments2D2012RfSpectraMolecules, 
Experiments2D2011Crossover2D3D, Experiments2D2012Crossover2D3D, Experiments2D2012Polarons, 
Experiments2D2011DensityDistributionTrapped,Experiments2D2012Viscosity, ContactExperiment2D2012, Vale2Dcriteria, 
Experiments2D2014, Experiments2D2015SpinImbalancedGas, Experiments2D2015PairCondensation, 
Experiments2D2015BKTObservation, Experiments2D2015Density}.

On the theory side, early work in 2D used mean-field approaches to
study the crossover between Bose-Einstein condensation 
(BEC) and Bardeen-Cooper-Schrieffer (BCS) pairing~\cite{Miyake,BCSBEC2D,ZhangLinDuan}. 
The ground-state energy and contact were computed in the thermodynamic limit in Ref.~\cite{Bertaina} using the diffusion 
Monte Carlo method, which was updated and expanded by Refs.~\cite{ShiChiesaZhang,GDGG} 
with a precise ab initio study of multiple ground-state properties. 
Studies at finite temperature have also appeared (see e.g.~\cite{LiuHuDrummond, Enss2D, ParishEtAl, BarthHofmann, ChaffinSchaefer, 
EnssUrban, BaurVogt, EnssShear, EoS2D,ValeComparison}).

Thus, a fair amount is known about the many-body physics of these systems;
however, much less is known about their few-body properties and how they approach the thermodynamic regime.
In this work, we calculate from first principles some of the
most important ground-state properties characterizing this few-to-many crossover: the energy, Tan's contact, the momentum
distribution, and the single-particle propagator.

\section{Hamiltonian and computational approach}
Since our intent is to focus on non-relativistic Fermi systems with short-range interactions,
our Hamiltonian is given by
\beq
\label{Eq:H}
\hat{H} = \hat{T} + g\,\hat V
\eeq
where the kinetic and potential terms $\hat{T}$ and $\hat{V}$ are given by
\beq
\hat{T} = \sum_{s=\uparrow,\downarrow}{\int{d^2 x\,\hat{\psi}^{\dagger}_{s}(\bd x)\left(-\frac{\hbar^2\nabla^2}{2m}\right)\hat{\psi}^{}_{s}(\bd x)}}
\eeq
and
\beq
\hat{V} = -\int{d^2 x\,\hat{n}^{}_{\uparrow}(\bd x)\hat{n}^{}_{\downarrow}(\bd x)},
\eeq
respectively, and where the spin-$s$ fermionic field operators are denoted with $\hat{\psi}^{}_{s}$ and $\hat{\psi}^{\dagger}_{s}$, 
with associated densities $\hat{n}^{}_{s}$. From this point on, we use units such that $\hbar = m = k_{\text B} = 1$, such that
$g$ is dimensionless and $\hat V$ has dimensions of energy as written. Although no new dimensionful parameters
enter the dynamics of the system when the interaction is turned on, the classical scale invariance is broken by quantum 
fluctuations, which result in a non-zero pair binding energy.

Given a stable many-body system, ground-state expectation values may be 
obtained from the large-imaginary-time properties of an arbitrary trial state $|\phi^{}_0\rangle$,
so long as the latter is not orthogonal to the system's true ground state. 
In this work, we take $|\phi^{}_0\rangle$ to be a single Slater determinant made out of the lowest-energy plane-wave 
orbitals. While this choice can certainly be optimized, we find it to be sufficient for our purposes.

For an operator $\hat O$, we define
\beq
O^{}_\beta \equiv \frac{\langle\phi^{}_0|
\,\hat{U}(\beta,\beta/2)\,\hat{O}\,\hat{U}(\beta/2,0)\,
|\phi^{}_0\rangle}{\langle\phi^{}_0|\,\hat{U}(\beta,0)\,|\phi^{}_0\rangle},
\eeq
where
\beq
\hat{U}(\tau^{}_b,\tau^{}_a)\equiv\exp\left[-(\tau^{}_b-\tau^{}_a)\hat{H}\right]
\eeq
is the imaginary-time evolution operator.  It follows immediately that
%
$O^{}_\beta \xrightarrow[]{\beta\to\infty}\langle\hat{O}\rangle$,
%
where the expectation value on the right is in the true ground state of the system.
For some observables, in particular the Hamiltonian itself, this convergence can be easily shown to be monotonic 
in $\beta$ (in fact, exponential), which makes their acquisition to some extent straightforward.

To address the interaction, we approximate the imaginary-time evolution operators $\hat{U}$ via a symmetric 
Suzuki-Trotter decomposition as
\beq
\hat{U}(\tau^{}_a+\tau,\tau^{}_a) = e^{-\tau\hat{T}/2}\;e^{-\tau g \hat{V}}\;e^{-\tau\hat{T}/2}+O(\tau^{3}),
\eeq
again splitting the Hamiltonian into its relatively simple one-body kinetic term and comparatively complicated two-body, zero-range potential term.
While continuous-time approaches have been known for a long time~\cite{RomboutsEtAl}, they have not yet been adapted to
the hybrid Monte Carlo technique, which we prefer in order to make contact with lattice-QCD methods~\cite{HMC}.
At each timestep, we decompose the central (potential energy, two-body) operator via a Hubbard-Stratonovich transformation~\cite{HS} 
into a linear combination of products of one-body operators writing (generically)
\beq
\label{Eq:HS}
e^{-\tau g \hat{V}} = \int{\mathcal{D}\sigma \; e^{-\tau\hat{V}^{}_{\uparrow,\sigma}}e^{-\tau\hat{V}^{}_{\downarrow,\sigma}}},
\eeq
for an auxiliary field $\sigma(\bd x)$ summed over all possible configurations at each imaginary-time slice. The
specific form of the operators $e^{-\tau\hat{V}^{}_{s,\sigma}}$, for $s=\uparrow,\downarrow$, depends on the 
choice of Hubbard-Stratonovich transformation. In our case, we have decoupled the interaction in the
density-density channel using a continuous and compact auxiliary field (see e.g. Ref.~\cite{QMCReviews} for
further details).

In the above, we have implicitly assumed that the number of spatial degrees of freedom at each time step is finite. We 
accomplish this by taking space to be a square lattice, which results in a lattice field theory approach in the same usual fashion 
as in lattice-QCD and Hubbard-model Monte Carlo calculations. The field integral of Eq.~(\ref{Eq:HS}) is estimated in practice
using Metropolis-algorithm-based Monte Carlo methods, which is possible as for unpolarized systems there is no sign problem.
Further details can be found in Ref.~\cite{QMCReviews}; closely related methods were used to examine 
systems in 1D in Ref.~\cite{EoS1D} and in 3D in Ref.~\cite{BDM1}.

In this work, we have used spatial lattice sizes of side $N_x^{}=24,28,32,36,40$ points, and taken the spatial lattice spacing to be $\ell = 1$ and
the temporal lattice spacing $\tau$ such that $\tau/\ell^2 = 0.01 - 0.05$.
While the method is not limited by these parameters, we found them sufficient to achieve the continuum limit and to
characterize the crossover from few- to many-body physics.
On the other hand, the Monte Carlo estimation of the field integrals carries a statistical uncertainty. To reduce the latter,
we took at least 500 decorrelated samples of the auxiliary field, such that the uncertainty can be expected to be of order $5\%$ or less.

\section{Results and Discussion}
To calibrate our lattice field theory, we solved the two-body problem for all values of $N_x^{}$ 
mentioned above and determined the lattice binding energy $\varepsilon_\text{B}^{}$ as a function of
the dimensionless coupling $g$ and the lattice size $N_x^{}$. Using those results, we performed
calculations for higher particle numbers $N=4,8,12,\dots,40$ at fixed physics as set by the renormalized coupling 
$\eta=1/2 \ln(2 \varepsilon_\text{F}/\varepsilon_\text{B})$, where 
$\varepsilon_\text{F}^{} =  k_\text{F}^{2}/2$ is the Fermi energy, $k_\text{F}^{} = \sqrt{2 \pi n}$ is the Fermi momentum,
and $n=N/L^2$ is the total density.
To ensure that our results are converged to the ground state, we followed the procedure outlined above 
of calculating at finite $\beta$ and extrapolating to $\beta \to \infty$.

\begin{figure}[t]
\includegraphics[width=1.0\columnwidth]{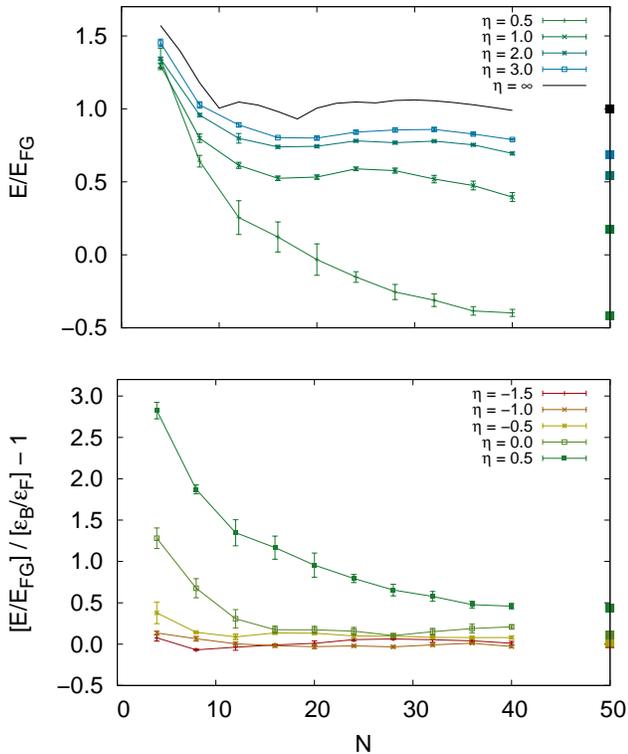}
\caption{\label{Fig:GSE}(color online) Ground-state energy $E$ of $N=4,8,12,...,40$ unpolarized fermions, 
for several values of the dimensionless coupling $\eta = -1.5,-1.0,\dots,3.0,\infty$, the final corresponding to a free system. 
Top panel shows $E$ for the four weakest couplings we calculated, 
in units of the energy of the noninteracting system $E^{}_\text{FG} = \frac{1}{2}N\varepsilon_\text{F}^{}$.
Bottom panel displays $E$ for the strongest couplings we considered, using
the binding energy per particle $\varepsilon^{}_\text{B}/2$ as a scale. 
Clearly, for $\eta < 0$ the energy per particle is dominated by the pair binding energy across all particle numbers.
In both plots, the ground-state energy results of Ref.~\cite{ShiChiesaZhang} in the thermodynamic
limit are shown with solid squares at $N=50$.
}
\end{figure}
\begin{figure}[b]
\includegraphics[width=1.0\columnwidth]{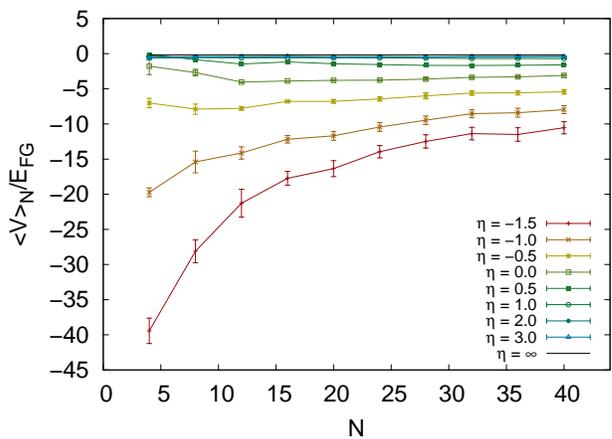}
\caption{\label{Fig:GSC}(color online) Expectation value of the interaction operator $\hat V$ for $N$=4,8,12,...,40 fermions, in units of the 
ground-state energy of the noninteracting gas, for several values of the dimensionless coupling $\eta = -1.5,-1.0,\dots,3.0$, along with the noninteracting case $\eta\to\infty$.
The solid line shows $-1/(2\pi)$, which is the result in the noninteracting limit.
}
\end{figure}

In Fig.~\ref{Fig:GSE}, we show our results for the energy (extrapolated to infinite volume), in units of the energy of the noninteracting case
$E^{}_\text{FG} = \frac{1}{2}N\varepsilon_\text{F}^{}$, as a function of particle number and coupling. For display
purposes, we have separated weak couplings (i.e. BCS side, shown in the top panel) from strong couplings 
(mostly BEC side, shown in the bottom panel).
As evident from the top panel, at weak couplings there is a structure of oscillations before the few-body problems
heal to the thermodynamic limit result (shown on the right-side of the figure with squares, using the
data of Ref.~\cite{ShiChiesaZhang} and interpolations thereof where needed).
Such oscillations are typically associated with so-called shell effects and have been seen in 
1D and 3D analogues of this system (see e.g.~\cite{GSC1D} and~\cite{ForbesGandolfiGezerlis}).
In order to better understand the strong coupling regime, we subtracted the binding energy per particle $\varepsilon^{}_\text{B}/2$ 
from the total energy per particle in the bottom panel. 
Indeed, for $\eta \leq 1.0$, the onset of the BEC regime implies that the energy is expected to be dominated by the
binding energy of the pairs, which form immediately upon turning on the interaction.
The numerical results plotted in Fig.~\ref{Fig:GSE} are given in Appendix~B, Table~\ref{Table:Energy}.

While the ground-state energy is an essential quantity in any few- and many-body problem, more detailed information about the
short-distance behavior of the system can be obtained from Tan's contact $C$~\cite{ContactReview}. Indeed, it has been shown that $C$
controls the high-momentum tail of the momentum distribution (see below)~\cite{TanContact}, as well as multiple sum rules 
of real-time response functions~\cite{ContactResponse}. The calculation of $C$ itself, however, involves a many-body 
problem that requires computational approaches~\cite{ContactNumerical,GSC1D}.
The contact obeys an adiabatic theorem (see~\cite{ContactAdiabatic,Valiente}), which indicates that $C$ is proportional to
the change in the ground-state energy $E$ with the s-wave scattering length $a_0^{}$. 
In our Hamiltonian~Eq.~(\ref{Eq:H}), the scattering length enters fully through the bare coupling $g$
(and, of course, the UV lattice cutoff, which we hold constant). Therefore,
\beq
C \propto \frac{\partial E}{\partial \ln (k^{}_\text{F} a^{}_0)} = \frac{\partial E}{\partial g}\frac{\partial g}{\partial \ln (k^{}_\text{F} a^{}_0)}.
\eeq
The factor ${\partial g}/{\partial \ln (k^{}_\text{F} a^{}_0)}$ is a derivative at constant $N$ that depends only on two-body physics, 
as the bare coupling $g$ is determined by tuning to the desired $a^{}_0$ by solving the two-body problem. The ${\partial E}/{\partial g}$ factor,
on the other hand, encodes many-body correlations and depends on the particle content $N$.
Because we have used a contact interaction [see Eq.~(\ref{Eq:H})], the expectation value of the potential energy gives us access to ${\partial E}/{\partial g}$
through the Hellmann-Feynman relation for the $N$-body problem:
\beq
\langle \hat V \rangle^{}_N = \frac{\partial E}{\partial g}.
\eeq
%
In Fig.~\ref{Fig:GSC}, we show $\langle \hat V \rangle^{}_N$ in units of the 
ground-state energy of the noninteracting gas $E^{}_\text{FG}$ and as a function of both particle number and coupling.
The numerical results plotted here are given in Appendix~B, Table~\ref{Table:Contact}.


In Fig.~\ref{Fig:nk}, we show our results for the momentum distribution as a function of $k/k_\text{F}^{}$ for $N=36$ particles
and for several values of the dimensionless coupling $\eta$, extrapolated to $\beta\to\infty$ (i.e. the ground state).
Data at finite $\beta$ are shown in Appendix A. 
The inset shows the same data in log-log form along with fits of the expected power law $k^{-4}$ (see e.g. Ref.~\cite{WernerCastin}),
which yield the contact at $k\gg k_\text{F}^{}$. The expected behavior is obtained at weak coupling, 
but the region where it is valid becomes increasingly limited (i.e. it moves toward high $k/k_\text{F}^{}$) at strong coupling. 
Thus, in order to see the expected momentum tail at strong coupling, calculations at larger volumes (lower $k^{}_\text{F}$) are needed.
For most of the couplings we studied, however, it appears that the $k^{-4}$ decay is reached around $k/k_\text{F}^{}\simeq 1.8-2.0$,
which is remarkably close to its 3D counterpart~\cite{ContactNumerical}.

\begin{figure}[t]
\includegraphics[width=1.0\columnwidth]{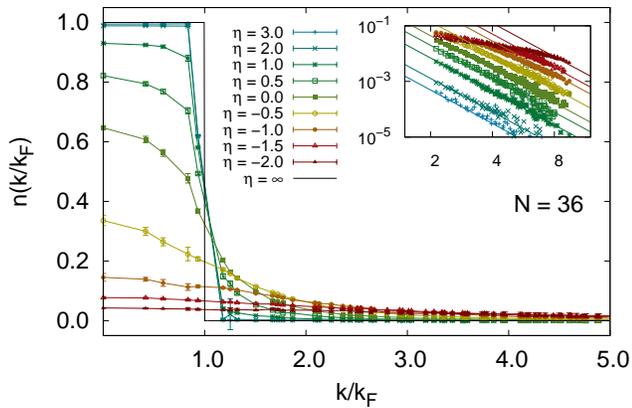}
\caption{\label{Fig:nk}(color online) Momentum distribution of $N=36$ unpolarized fermions, as a function of $k/k_\text{F}^{}$, for several 
values of the dimensionless coupling $\eta = -2.0,-1.5,...,3.0$, as well as the noninteracting case. Inset: Momentum distribution in log-log scale, showing the power-law
decay that heals to a $\sim (k/k_\text{F}^{})^{-4}$ decrease at large $k$.
}
\end{figure}
\begin{figure}[b]
\includegraphics[width=1.0\columnwidth]{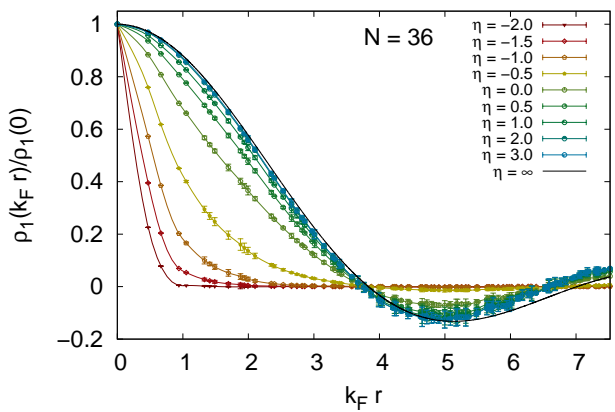}
\caption{\label{Fig:GF1}(color online) One-body density matrix $\rho^{}_1$ for $N=36$ unpolarized fermions, as a function of $k_\text{F}^{} r$, for several 
values of the dimensionless coupling $\eta = -2.0,...,3.0$, along with the $\eta\to\infty$ (i.e. noninteracting) case. 
}
\end{figure}
%
In Fig.~\ref{Fig:GF1}, we show our results for the one-body density matrix $\rho^{}_1$ defined in our
unpolarized system as
\beq
\rho^{}_1(\bd x,\bd x') = \langle \hat{\psi}^{\dagger}_{\uparrow}(\bd x)\hat{\psi}^{}_{\uparrow}(\bd x')\rangle = 
\langle \hat{\psi}^{\dagger}_{\downarrow}(\bd x)\hat{\psi}^{}_{\downarrow}(\bd x')\rangle,
\eeq
given the figure as a function of the dimensionless distance $k_\text{F}^{} r$, where $r = |\bd x-\bd x'|$, as we take into account 
translation and rotation invariance.
The results shown are for $N=36$ particles and cover several values of the coupling $\eta$. 
The localized shape of $\rho^{}_1$ around $x=0$ at strong couplings
is a direct manifestation of the formation of bound pairs, which in turn makes lattice approaches to the problem more challenging:
the presence of the lattice-spacing scale competes with the pair size, which must be properly resolved in order to obtain accurate results.
 
To encode the intermediate and short-distance ($k^{}_F r < 3.0$) shape of our numerical results for $\rho^{}_1$, we fit the following 
dimensionless form to the data of Fig.~\ref{Fig:GF1}:
\beq
\label{Eq:FIT}
f(k_F^{} r,\eta) = 2 \, e^{- a\, p^{}_{\text B} r}  \, \frac{J_1(k_{\text F}^{}r)}{k_{\text F}^{} r},
\eeq
where $a$ is a fit parameter used to interpolate across couplings, and $p^{}_{\text B} = \sqrt{2}k^{}_{\text F} e^{-\eta}$ is the binding momentum of the two-body system (see Table~\ref{TableABEta} for fit results).
The exponential factor in Eq.~(\ref{Eq:FIT}) is motivated by the deep bound state in the BEC regime, where single-particle correlation lengths 
are expected to be governed by the inverse binding momentum.
The Bessel function factor, along with the denominator, corresponds to the noninteracting case in the continuum limit.(see Table I for fit results)

\begin{table}[h]
\begin{center}
\caption{\label{TableABEta}
Fit parameter $a$ obtained by fitting Eq.~(\ref{Eq:FIT}) to the data of Fig.~\ref{Fig:GF1},
as a function of the dimensionless coupling $\eta$. The rightmost column shows the 
sum of the absolute value of the residuals per degree of freedom.
These values of $a$ exemplify the typical numbers obtained across all particle numbers, 
i.e. beyond the data of Fig.~\ref{Fig:GF1}.
}
\begin{tabularx}{\columnwidth}{@{\extracolsep{\fill}}c c c}
\hline \hline
$ \eta $ 	&  $a$		& 	Mean residual\\
\hline
$-2.0$   	& $0.45(3)$	&	$0.001$ \\
$-1.5$ 	& $0.44(3)$	&	$0.002$ \\
$-1.0$   	& $0.42(3)$	&	$0.004$ \\
$-0.5$ 	& $0.27(6)$	&	$0.03$ \\
$ 0.0$    	& $0.11(4)$	&	$0.02$ \\
$ 0.5$  	& $0.06(4)$	&	$0.02$ \\
$ 1.0$    	& $0.05(7)$	&	$0.02$ \\
$ 2.0$    	& $0.0(1)$		&	$0.02$ \\
$ 3.0$    	& $0.0(1)$		&	$0.02$ \\
\hline  \hline
\end{tabularx}
\end{center}
\end{table}

\section{Summary and Conclusions}
In this work, we set out to examine, in a fully non-perturbative fashion, the progression from
few- to many-body fermions in 2D, with attractive short-range interactions across a wide range of coupling strengths in
the BEC-BCS crossover.
Using lattice Monte Carlo methods akin to those of lattice QCD, we calculated the universal 
behavior of systems of $N=4,8,...,40$ particles in the ground state. We focused, in particular, on the energy 
per particle, Tan's contact, the momentum distribution, and the single-particle correlator.
The ground-state energy for each coupling strength forms a smooth curve with mild oscillations toward
the thermodynamic limit.
For $\eta \leq 0$, particularly for $N > 16$, the ground-state energy
is completely dominated by the binding energy of the pairs. Thus, the thermodynamic limit is reached
much faster on the BEC side than on the BCS side.
The momentum distribution approaches a $(k/k_\text{F}^{})^{-4}$ decay at large $k$, as expected, but 
the onset of that behavior shifts noticeably towards large $k$ as the coupling is increased.
Finally, our fits to the single-particle correlation function $\rho^{}_1$ indicate a shift from a $k_F^{}$-dominated region
at weak coupling, to a $p_\text{B}^{}$-dominated region at strong coupling.

\acknowledgments
{
We gratefully acknowledge discussions with E. R. Anderson and J. Kaufmann.
This material is based upon work supported by the National Science Foundation under 
Grants 
No. PHY1306520 (Nuclear Theory program) 
and 
No. PHY1452635 (Computational Physics program).
}

\section{Appendix A: Finite-volume effects and extrapolations}
\label{app:a}

\begin{figure}[b]
\includegraphics[width=1.0\columnwidth]{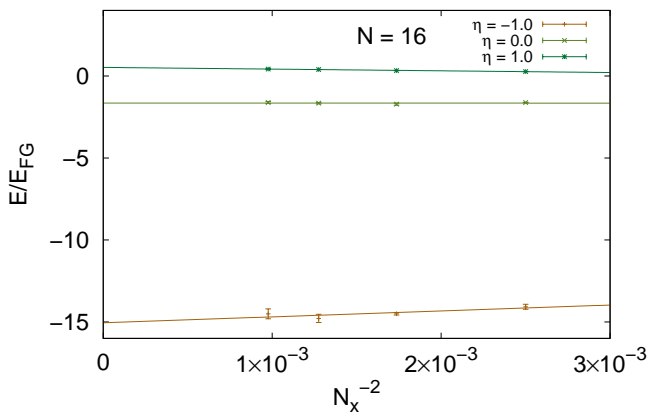}
\caption{\label{Fig:volumeFixedN}(color online) Extrapolation to the infinite volume limit for $N = 16$ unpolarized fermions, as a function of $N_x^{-2}$, for several values of the dimensionless coupling $\eta = -1.0, 0.0, 1.0$.}
\end{figure}

\begin{figure}[t]
\includegraphics[width=1.0\columnwidth]{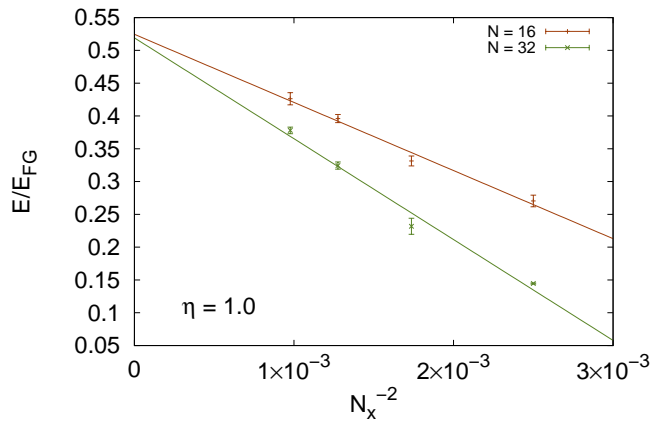}
\caption{\label{Fig:volumeFixedETA}(color online) Extrapolation to the infinite volume limit for $N = 16,32$ unpolarized fermions at fixed dimensionless coupling $\eta = 1.0$, as a function of $N_x^{-2}$. }
\end{figure}

In this appendix we elaborate on the procedure we employed to carry out extrapolations to the
infinite volume limit, and present specific examples.
In Fig.~\ref{Fig:volumeFixedN} we show the extrapolation for $N=16$ particles for three different values of
the dimensionless coupling $\eta$. To complement that example, we show in Fig.~\ref{Fig:volumeFixedETA}
the infinite-volume extrapolation for two different particle numbers at a fixed coupling.


\begin{figure}[b]
\includegraphics[width=1.0\columnwidth]{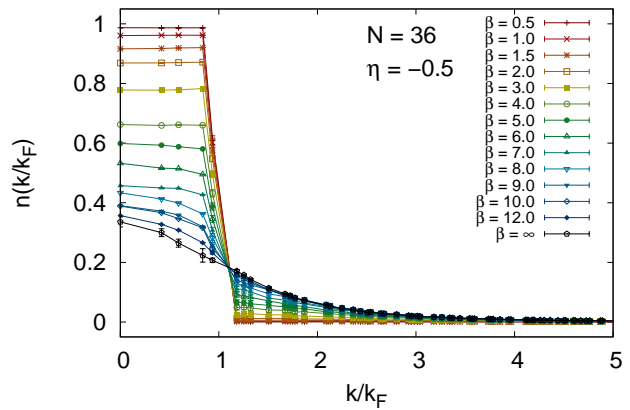}
\caption{\label{Fig:nkETADecay}(color online). Momentum distribution extrapolated to infinite imaginary time $\beta$ for $N = 36$ unpolarized fermions on a $32 \times 32$ lattice  and at fixed dimensionless coupling $\eta = -0.5$. }
\end{figure}

In Fig.~\ref{Fig:nkETADecay} we show, for a specific coupling $\eta = -0.5$, the results of extrapolating our data for the momentum
distribution $n(k)$ to the ground state, which we accomplish by increasing the length $\beta$ of the imaginary time direction 
(i.e. the projection time). Clearly, such a coupling, though intermediate in strength, requires $\beta \simeq 8.0$ to start converging to 
the large $\beta$ limit, in particular for the region $k < k^{}_F$. One way to overcome such large projection times is to use a different
guess for the ground-state wavefunction, as done for instance in Refs.~\cite{ShiChiesaZhang,GDGG}.


\section{Appendix B: Ground-state data tables}
\label{app:b}

In this appendix we present tables showing our estimates for the ground-state energy
and Tan's contact as a function of the coupling $\eta=-1.5,-1.0,...,3.0$, for particle numbers $N=4, 8, 12,\dots, 40$,
extrapolated to the infinite-volume limit.

\onecolumngrid

%
\begin{table}[h]
\begin{center}
\caption{\label{Table:Energy}
Ground-state energy $E$ on the lattice, in units of the continuum noninteracting energy $E^{}_\text{FG}= N \varepsilon_F^{}/2$ of the 
$N$-particle system, as a function of $N$ and the dimensionless coupling $\eta$.}
\begin{tabularx}{\columnwidth}{@{\extracolsep{\fill}}c c c c c c c c c c}
\hline \hline
\multicolumn{10}{c}{$ \eta $} \\
\cline{2-10}
$N$	& -1.5	& -1.0	& -0.5	& 0.0	& 0.5	& 1.0	& 2.0	& 3.0 & $\infty$ \\
\hline
4	& -37(2)	& -12.8(3)	& -3.4(7)	& 0.6(2)	& 1.34(7)	& 1.30(2)	& 1.34(1)	& 1.45(3)	& $\pi/2$	\\
8	& -43.0(3)	& -13.8(4)	& -4.66(5)	& -0.6(2)	& 0.64(4)	& 0.80(3)	& 0.96(1)	& 1.03(2)	& $3\pi/8$	\\
12	& -41(2)	& -14.7(3)	& -5.0(2)	& -1.4(2)	& 0.3(1)	& 0.61(2)	& 0.80(3)	& 0.89(1)	& $4\pi/12$	\\
16	& -40.6(2)	& -15.0(3)	& -4.70(4)	& -1.65(9)	& 0.1(1)	& 0.52(2)	& 0.74(1)	& 0.80(0)	& $5\pi/16$	\\
20	& -39(1)	& -15.2(4)	& -4.72(6)	& -1.65(9)	& -0.0(1)	& 0.53(2)	& 0.74(1)	& 0.80(1)	& $8\pi/25$	\\
24	& -38.0(8)	& -15.1(2)	& -4.9(1)	& -1.69(9)	& -0.15(4)	& 0.59(1)	& 0.78(0)	& 0.84(1)	& $4\pi/12$	\\
28	& -37.6(3)	& -15.3(2)	& -4.91(8)	& -1.79(4)	& -0.25(5)	& 0.58(2)	& 0.77(1)	& 0.86(1)	& $33\pi/98$	\\
32	& -38(1)	& -14.9(3)	& -4.99(1)	& -1.70(8)	& -0.31(4)	& 0.52(3)	& 0.78(1)	& 0.86(1)	& $43\pi/128$	\\
36	& -38.5(6)	& -14.6(1)	& -5.01(7)	& -1.6(1)	& -0.38(3)	& 0.48(3)	& 0.75(1)	& 0.82(0)	& $53\pi/162$	\\
40	& -39.8(9)	& -15.2(2)	& -5.00(9)	& -1.58(4)	& -0.40(2)	& 0.40(3)	& 0.70(0)	& 0.80(1)	& $63\pi/200$	\\
\hline  \hline
\end{tabularx}

\end{center}
\end{table}
%

\begin{table}[h]
\begin{center}
\caption{\label{Table:Contact}
Ground-state interaction $\langle \hat V \rangle^{}_N$, in units of the energy of the non interacting gas $E^{}_\text{FG}$, 
as a function of $N$ and the dimensionless coupling $\eta$.
}
\begin{tabularx}{\columnwidth}{@{\extracolsep{\fill}}c c c c c c c c c c}
\hline \hline
\multicolumn{10}{c}{$ \eta $} \\
\cline{2-10}
$N$	& -1.5	& -1.0	& -0.5	& 0.0	& 0.5	& 1.0	& 2.0	& 3.0	& $\infty$ \\
\hline
4	& -39(2)	& -19.7(6)	& -7.0(7)	& -1(1)	& -0.1(2)	& -0.34(5)	& -0.50(1)	& -0.6(2)	 &-1/(2$\pi$)	\\
8	& -28(2)	& -15(1.5)	& -7.9(7)	& -2.7(5)	& -0.86(9)	& -0.55(9)	& -0.41(1)	& -0.5(1)	 &-1/(2$\pi$)	\\
12	& -21(2)	& -14.1(9)	& -7.8(3)	& -4.0(1)	& -1.47(4)	& -0.59(1)	& -0.42(4)	& -0.44(7)	 &-1/(2$\pi$)	\\
16	& -17.7(1)	& -12.2(5)	& -6.8(1)	& -3.9(1)	& -1.2(2)	& -0.58(1)	& -0.38(2)	& -0.43(2)	 &-1/(2$\pi$)	\\
20	& -16(1)	& -11.7(6)	& -6.8(3)	& -3.8(1)	& -1.44(8)	& -0.59(4)	& -0.41(2)	& -0.52(6)	 &-1/(2$\pi$)	\\
24	& -14.0(9)	& -10.4(6)	& -6.4(3)	& -3.8(1)	& -1.55(1)	& -0.60(1)	& -0.41(1)	& -0.50(5)	 &-1/(2$\pi$)	\\
28	& -12.5(9)	& -9.5(6)	& -6.0(4)	& -3.6(2)	& -1.67(3)	& -0.64(1)	& -0.44(1)	& -0.48(5)	 &-1/(2$\pi$)	\\
32	& -11.4(9)	& -8.5(6)	& -5.6(3)	& -3.4(1)	& -1.68(1)	& -0.69(4)	& -0.41(1)	& -0.46(4)	 &-1/(2$\pi$)	\\
36	& -11.5(1)	& -8.4(6)	& -5.6(3)	& -3.3(1)	& -1.65(1)	& -0.68(3)	& -0.40(2)	& -0.45(4)	 &-1/(2$\pi$)	\\
40	& -10.5(9)	& -8.0(6)	& -5.4(3)	& -3.1(1)	& -1.59(2)	& -0.71(3)	& -0.41(2)	& -0.43(3)	 &-1/(2$\pi$)	\\
\hline  \hline
\end{tabularx}

\end{center}
\end{table}
%

\twocolumngrid


\end{document}